# AI and the future of academic peer review

**Sebastian Porsdam Mann,[1,2] Mateo Aboy,[3] Joel Jiehao Seah,[2] Zhicheng Lin,[4] Xufei Luo,[5,6] Daniel Rodger,[7] Hazem Zohny,[8] Timo Minssen,[1] Julian Savulescu,[2,8*] and Brian D. Earp[2]**

**1. Center for Advanced Studies in Bioscience Innovation Law (CeBIL), Faculty of Law, University of Copenhagen**

**2. Center for Biomedical Ethics, Yong Loo Lin School of Medicine, National University of Singapore**

**3. Faculty of Law, University of Cambridge**

**4. Department of Psychology, Yonsei University, Seoul**

**5. Vincent V.C. Woo Chinese Medicine Clinical Research Institute, School of Chinese Medicine, Hong Kong Baptist University**

**6. Chinese EQUATOR Centre, Hong Kong**

**7. School of Allied and Community Health, London South Bank University, London**

**8. Uehiro Oxford Institute, University of Oxford, Oxford**

*** Corresponding author.**

## Abstract

Pre-publication peer review remains, aspirationally, a central quality-control mechanism of science and scholarship. Yet its ability to fulfill this role is increasingly strained: long publication delays, escalating reviewer burden concentrated on a small minority of scholars, inconsistent quality and low inter-reviewer agreement, and systematic biases by gender, language, and institutional prestige persist. Decades of human-centered reforms have yielded only marginal improvements. Meanwhile, Artificial Intelligence (AI) systems—especially large language models (LLMs) and agentic systems based on these—are being piloted across the peer-review pipeline by journals, publishers, funders, and individual reviewers. Early studies suggest that AI assistance can produce reviews comparable in quality to human reviews, accelerate reviewer selection and feedback, and reduce certain biases. But it also raises distinctive concerns about ontology, hallucination, novelty recognition, human oversight, confidentiality, quality science, gaming, and trust. In this paper, we map the aims and weaknesses of peer review to specific LLM applications that might help to address such weaknesses, while also dealing with objections and pointing to safeguards that could make their use acceptable. We show that targeted—and appropriately human-supervised—LLM assistance can likely improve error detection, timeliness, and reviewer workload concerns without displacing human judgment. We highlight advanced architectures, including fine-tuned, retrieval-augmented, and multi-agent systems, that may enable more reliable, auditable, and interdisciplinary review. We argue that much depends on specific ethical and practical considerations: the legitimacy of AI-assisted peer review will turn on governance choices as much as technical capacity. Between the Scylla of uncritical adoption and the Charybdis of reflexive rejection, we argue for carefully scoped pilots with explicit evaluation metrics, transparency through disclosure, author choice, and human judgment and accountability.

## Introduction

Academic peer review serves as an important quality control mechanism for scholarly research. Although post-publication approaches are increasingly discussed, peer review is typically carried out prior to formal publication with the aim of providing a fair and impartial recommendation about whether a given contribution should be published. The process relies on expert judgment—traditionally that of human subject-matter specialists recruited on an *ad hoc* basis by the handling editor(s) of an academic journal—to: assess originality, clarity, and coherence; evaluate methods; detect errors; and suggest improvements to submitted manuscripts.

Yet the system faces well-documented challenges. These include overwhelming submission volumes coupled with a limited pool of competent reviewers (who contribute voluntarily, and are usually unpaid), decreasing willingness to review (Bravo et al., 2019), persistent biases against certain groups and institutions (Murray et al., 2018), inconsistent review quality (Bornmann et al., 2010), and lengthy delays that impede scholarly and scientific progress (Andersen et al., 2021). Previous attempts to address these problems through human-centered interventions, such as reviewer training programs, open review systems, and quality checklists, have yielded disappointing results, suggesting that incremental reforms to the existing system may be insufficient (Bruce et al., 2016; Gaudino et al., 2021).

Artificial intelligence (AI) approaches, including generative AI systems based on large language models (LLMs), are now being piloted across the peer review pipeline. Major publishers, including Springer Nature, Elsevier, and Wiley, are experimenting with AI-assisted tools at different points in the publication workflow, including integrity screening, reviewer matching, and editorial quality checks. Scientific societies, like AAAI, are testing LLM-integrations in supplementing the initial reviews of conference paper submissions. Researchers have started comparing human and AI assistance in review quality and in detecting AI-mediated writing (Shcherbiak et al., 2024; Liang et al., 2024). And funders like the U.S. NIH have begun issuing policies on the use of AI in grant applications and peer review.

While commentators are beginning to weigh in on these developments (e.g. Hyde, 2026), the pace of experimentation and technological advancement is moving at breakneck speed. As of writing, the scientific community lacks clear and comprehensive guidance on when and how to use LLMs, what safeguards are necessary, and where benefits plausibly outweigh risks. The purpose of this paper is to provide a pragmatic, ethics-informed decision map: a linkage from (i) the aims and weaknesses of peer review, to (ii) concrete LLM applications that might help to address these weaknesses and promote said aims, to (iii) the principal objections such applications raise, and (iv) the safeguards and evaluation criteria that would render their use acceptable. We offer an overview of the evidence and arguments both in favor of, and against, the use of AI in peer review. We suggest it is an open question whether AI-assisted peer review will prove beneficial, all things considered. Much will depend on implementation choices—such as editorial policies, reviewer training, and system governance—in addition to technical capacity. However, we argue that institutional responses oriented around reflexive prohibition or indefinite deferral of AI in peer review carry costs that must also be weighed. Where evidence of benefit exists, failing to act on it (with all due caution and restraint) is not an ethically neutral position.

## What is peer review supposed to accomplish—and is it succeeding?

Peer review serves multiple interconnected functions that, when realised, collectively serve to uphold the integrity of scholarly and scientific knowledge production, thereby contributing to the wider public's (and policymakers') trust in the credibility of published research (Kharasch et al., 2021). At its core, the process exists to assess the originality and significance of a contribution, evaluate the rigor of its methods and reasoning, and determine whether the conclusions are adequately supported by the evidence—all while adhering to the norms of fairness, timeliness, and constructive engagement that define scientific practice at its best (Turner, 2003). The system is also supposed to act as a safeguard against error and misconduct, from catching statistical mistakes and logical fallacies to detecting plagiarism, data fabrication and manipulation, and undisclosed conflicts of interest (COIs) (but see Stroebe et al., 2012). However, despite these aspirations, there are well-known and well-documented concerns around (a lack of) replicability, validity, and generalizability in many areas of science and medicine—the so-called "replication crisis"—often attributed to a combination of "questionable research practices," misaligned incentives between career progression and truth-seeking, and an unknown amount of hard-to-detect fraud (Mede et al., 2021; Nelson et al., 2022; Brown et al., 2022; Wible, 2022; Korbmacher et al., 2023). While there is no "quick fix" to these problems, there is clearly room for improvement in the extent to which peer review can be relied upon as a pre-publication filter for low-quality contributions.

Some authors have suggested that AI could play a role, in the future, in addressing at least some of the causes of the replication crisis (e.g., by detecting statistical errors in submitted manuscripts, or flagging potential author COIs), or, alternatively, in identifying already-published studies that have a low likelihood of being reproducible (e.g., based on pattern-matching with previous studies that have faced reproducibility challenges) (Straiton, 2024). Others, however, caution that AI itself—or rather, the use of AI by researchers who do not adequately scrutinise its outputs—could just as well serve to exacerbate the crisis at scale, potentially making it considerably worse (e.g., Ball, 2023; Al-leimon & Juweid, 2025).

The stakes of effective peer review thus extend far beyond individual manuscripts to encompass the credibility of entire research programs, fields of study, or even all of science. This, in turn, has implications for policy decisions grounded in scientific evidence. At the same time, publication decisions based on reviewer recommendations also determine career trajectories through their impact on hiring, promotion, and grant funding. This gatekeeping function generates inherent tensions: reviewers must balance thoroughness against efficiency, maintain standards while encouraging innovation, and provide critical assessment without discouraging legitimate challenges to orthodoxy. It is unsurprising that the system is buckling under pressures that existing reforms have not resolved.

## Specific problems with peer review

The ability of peer review to fulfil its functions is inhibited along several key dimensions.

First, the progress of scholarship depends on the availability of recent information, and thus on the rapid dissemination of scholarly works. This is true for all disciplines, but delays are especially consequential in certain fast-moving fields, such as AI or pandemic response. Yet, across fields, the journey from submission to publication often lasts many months, sometimes stretching into years. A cross-field analysis found typical end-to-end lags of roughly 9–18 months, varying by discipline and venue (Björk & Solomon, 2013). In line with these results, a more recent systematic review found large variations also within fields, with median submission-to-publication times for biomedical manuscripts ranging between 70 and 558 days (Andersen et al., 2021). These delays slow the dissemination of potentially useful results that have undergone evaluation by experts. They can also disproportionately burden early-career researchers, often employed on time-limited grants rather than in tenured or tenure-track positions, and may encourage questionable or inappropriate "strategic" behaviours (e.g., uninvited resubmissions, or multiple simultaneous submissions) that are not conducive to epistemic progress. They can also negatively impact graduate students in terms of graduation requirements, career opportunities, and mental and physical health in "publish or perish" academia (Friedrich et al., 2023; Wu, 2025). The practice of posting preprints (i.e., manuscripts made publicly available before having undergone formal peer review) might mitigate some of these harms, but they are an imperfect solution. First, their validity and value is often seen as uncertain, insofar as they have not yet been evaluated by independent experts, and hence, they may not be counted in evaluations, promotion decisions, and so on. Second, some journals still restrict or prohibit authors from posting submitted manuscripts onto a preprint repository.

Another concern is the sheer economic costs of peer review. Measured in time and foregone salary, reviewer effort amounts to what is essentially a very large subsidy (for journal publishers, other scholars, and wider society). A recent estimate placed global reviewer time in 2020 at well over 100 million hours, with the imputed wage value for the United States alone exceeding US$1.5 billion (Aczel et al., 2021). A separate survey estimated the annual global economic value of peer review at approximately US$6 billion when including reviewed-but-rejected manuscripts (LeBlanc et al., 2023). These burdens are not evenly distributed; multiple datasets show a highly skewed workload in which a small minority of scholars handles a large share of reviews. According to Publons (2018), which runs a service aimed at recognizing peer reviewers, approximately 10% of reviewers account for around 50% of all reviews; early-career researchers are disproportionately more likely than their senior colleagues to accept reviews (Bravo et al., 2019).

Since journal peer review is typically unpaid and anonymous, and seldom recognized, and academic advancement remains intensely competitive, individual incentives to conduct reviews are weak: reviewers overwhelmingly cite time pressure as the main reason to decline; most want formal recognition from institutions; and very few attach their names to reviews (for example, to avoid retaliation in case of a negative evaluation), which limits reputational returns (Bravo et al., 2019; Publons, 2018; Fox, 2021). The result is a classic collective action problem: it is in everyone's interests that peer review invitations are widely and promptly accepted, but it is not in most researchers' individual-level interests to conduct reviews. Thus, we have a situation in which conscientious (often junior) scholars absorb a disproportionate share of labour, resulting in unfair opportunity costs, while the system also moves more slowly than it might in the absence of these issues (Bravo et al.; 2019).

Given such mis-aligned incentives, combined with the ever-increasing volume of manuscript submissions and an ecosystem defined by the seemingly constant proliferation of new (often low-quality or "predatory") journals (Richardson et al., 2025), it is unsurprising that refusals of peer review invitations have increased over recent years. In a study of five Elsevier journals, the average percentage of accepted invitations to peer review declined from 43.6% to 30.9% between 2010 and 2017 (Bravo et al., 2019). Partly as a result of both this and the increasing number of submissions over time, total peer review invitations have grown by an estimated 9.8% per year (Publons, 2018). Fewer acceptances per invitation means longer handling times from having to search for other willing and suitable reviewers (who may be juniors in their fields), and heavier reliance on the same few names, perpetuating the skew in who bears the burden.

Once a peer review invitation is accepted, a different set of issues arises. Most journals require, or at least prefer, more than one reviewer; this is, in theory, a way to decrease the chance of bias, increase the odds of spotting errors, and help ensure the necessary expertise reviews interdisciplinary submissions. However, inter-reviewer agreement for journal manuscripts across studied fields is routinely low (mean Cohen's Kappa, a standard measure of agreement between raters, $\approx 0.17$) (Bornmann et al., 2010). In internal medicine, for example, reviewers' accept/revise/reject recommendations showed agreement "barely beyond chance" (Kravitz et al., 2010). Thus, the outcome of peer review may have as much to do with chance as with the quality of the manuscript assessed.

Further, consistent with our observations above, there is evidence that peer review only catches a small percentage of errors in manuscripts. Randomized "sting" experiments that deliberately seeded manuscripts with serious issues showed that reviewers missed many of them. In a landmark BMJ/JAMA study, reviewers detected on average only a quarter (two of eight) of injected design or analysis weaknesses; only 10% of reviewers spotted four or more, and 16% detected none (Godlee et al., 1998). A subsequent study also found that reviewers caught less than a third of deliberately inserted major errors (Schroter et al., 2008); and training programs targeted at improving these detection rates showed only minor improvements not sustained past a six-month follow-up (Schroter et al., 2004).

These issues are compounded by multiple intersecting biases. For example, editors tend to choose reviewers of their own gender and from familiar networks (Murray et al., 2018). Because men occupy a larger share of editorial roles, male editors' homophily amplifies male over-representation among reviewers; female editors also show same-gender preference, though typically weaker (Murray et al., 2018). Similarly, in a study looking at the single-blind review system (in which the identities and affiliations of submitting authors are made available to peer reviewers) at a high-end computer science conference, very large biases towards recommending acceptance for papers from famous authors, prestigious universities, and prestigious private enterprises were found (resulting in relative odds ratios of 1.63, 1.58, and 2.10, respectively) (Tomkins et al., 2017). Experimental work further suggests that, consistent with linguistic bias, identical scientific content written in non-native-like English may receive lower quality ratings (Politzer-Ahles et al., 2020).

However, none of this implies that peer review is bereft of any benefits (Smith, 2006). Especially when done well, it *can* result in major improvements to certain manuscripts, and *some* errors and weaknesses *are* caught by peer reviewers prior to publication. But the empirical record shows that the system, as practiced, often falls significantly short of its own ideals. It is subject to bias; it misallocates hidden labour; and it slows the diffusion of useful scholarship without always adding commensurate epistemic value. For our purposes, the upshot of all this is the following: any proposal to deploy AI-assisted/-based tools in peer review must be judged not against a romanticized ideal, but against the substantially flawed reality of the status quo.

**How AI can assist peer review**

The use of automated tools in peer review has been explored and employed (in certain limited forms) for decades. Early applications focused on specific tasks: plagiarism detection through systems like Turnitin (founded in 1998) and CrossRef's Similarity Check (introduced in 2008), which created databases of millions of articles for manuscript screening. The Toronto Paper Matching System introduced in 2010 used topic modeling to match submissions with reviewers and was adopted by leading computer science conferences (Charlin & Zemel, 2013). Statistical verification tools like Statcheck have been employed since 2015 to detect reporting errors (Nuijten et al., 2016), while the Granularity-Related Inconsistency of Means (GRIM) test has been used since 2016 to identify inconsistent means in approximately 50% of tested psychology articles (Brown & Heathers, 2017). While these systems reduced editorial workload, they remained limited to narrow tasks and could not fully replicate the evaluative functions of human review (Kousha & Thelwall, 2024).

These systems predate modern AI tools like LLMs. On a theoretical level, several characteristics of LLMs suggest potential value for peer review. They can process large volumes of data quickly, potentially identifying errors or relevant prior work that reviewers might miss. They can provide immediate feedback over multiple rounds. Their multilingual capabilities could help reduce the difference in prose quality between native and non-native English speakers. Several LLMs could be asked to collaboratively review interdisciplinary work from each relevant disciplinary perspective. Perhaps most fundamentally, they could help to address the reviewer shortage (through various mechanisms described below).

Yet there are reasons to worry about the (mis)use of LLMs in peer review. Generative AI systems can 'hallucinate,' generating fabricated references or errors not present in manuscripts (especially if forced to rely on their parametric knowledge rather than being given access to appropriate tools such as reference databases) (Donker, 2023; Huang et al., 2025). They risk reproducing training data biases or dominant assumptions, potentially suppressing innovative approaches (Hosseini & Horbach, 2023). Moreover, they could potentially compromise the security of manuscripts containing confidential research insofar as the latter are uploaded to commercial AI systems (NIH, 2023). In addition, ethical concerns about transparency have been raised, including: questions about how AI systems are developed, trained, and used; and calls to disclose details such as the system version, prompts, and date of use when AI is involved in the review process (Naddaf, 2025). Finally, LLMs are vulnerable to "prompt injection" and other adversarial attacks where hidden instructions manipulate outputs (Collu et al., 2025; Ye et al., 2024). We address these and other issues in the following section.

*Evidence on general-purpose LLMs in peer review*

The evidence on LLMs in peer review is early, heterogeneous, and context-specific (venues, tasks, prompts, and models vary). We therefore confine our claims to defined tasks in specific contexts, and treat any general conclusions as low-certainty signals that support pilot evaluation but not broad proof of superiority (or even effectiveness or acceptability, as the case may be).

With that caveat in mind, there is mounting, indicative evidence that reviewers and researchers already find value in AI for peer review assistance. Some peer reviewers already use LLMs to draft or polish reviews, particularly in computer science conferences (Naddaf, 2025; Latona et al., 2024). While self-reported uptake demonstrates a willingness to employ these tools for certain types of uses, the stronger question is whether such uses improve review quality without adding overriding new problems.

Emerging empirical work has begun to address this. One set of studies examined overall quality of peer review, as assessed by independent experts or by the authors receiving the review(s). These generally found that LLM-generated feedback can be useful to many researchers and can overlap with human review comments at rates similar to inter-reviewer overlap. For example, the International Conference on Learning Representation (ICLR) found that offering optional LLM feedback to randomly selected reviewers (on a draft version of their review) improved the informativeness and length of revised reviews, as assessed by blinded raters (Thakkar et al., 2025). A large study of more than 5,000 papers from Nature journals, ICLR, and eLife (Liang et al., 2024) reported that 57.4% of researchers in their sample found blinded reviews from GPT-4 (now a retired model) to be helpful overall, and 82.4% rated these as more helpful than at least some human reviewers.

There is also evidence that LLM assistance with (certain aspects of) peer review can help to address multiple interlocking concerns that plague the existing system. For one thing, the judicious use of LLMs has the potential to reduce *unnecessary latency* across stages of the peer review process. For example, LLMs have been shown, in some settings, to shorten reviewer-matching time (e.g., one study reported a 73% reduction in time-to-identify suitable reviewers (Farber, 2024)). As far as we are aware, whether they change time-to-first-decision or submission-to-publication remains untested. However, even moderate improvements, if not outweighed by other considerations, could substantially help with several related issues, including the increasing number of peer review invitations (since invitations might be better targeted; see below), the substantial economic burden of peer review (if reviewers can produce quality reports more efficiently), as well as, of course, the substantial delays in publishing (by compressing turnaround times at multiple editorial stages). Moreover, LLM assistance is easy to scale, thus partially addressing review issues related to the growth of scientific publishing.

The use of AI systems also addresses the disproportionate burden of peer review on unpaid volunteers and junior scholars. They do this indirectly, by reducing the overall burden; and may also do so directly, for example by helping to identify a broader pool of potential reviewers (cf. Stelmakh et al., 2021), better matching of these reviewers to papers, and (more speculatively) by helping to keep track of reviewer workloads and scheduling of requests (Zhang et al., 2025).

What about consistency, that is, agreement between reviewers? So far, there appears to be only one study that directly addresses this question, and the evidence remains preliminary. Liang et al. (2024) found that GPT-4's feedback overlaps with that of an individual human reviewer by approximately 30.9% for Nature-journal submissions (versus 28.6% human-to-human overlap), and by 39.2% for the ICLR dataset (versus 35.3% human-to-human overlap). This suggests that AI–human agreement is comparable to, or even better than, human–human agreement.

Similarly, preliminary studies suggest that LLMs can identify planted or known manuscript errors. Liu and Shah (2023) inserted deliberate mathematical and conceptual mistakes into 13 short computer science papers and found that GPT-4 was able to identify errors in 7 of them, suggesting some promise but far from comprehensive reliability. Another recent study (Zhang and Abernethy, 2025) used papers withdrawn from arXiv with known critical errors (whether factual, methodological, or logical flaws) to evaluate how well LLM-based quality checkers detect such errors. The best performing model in the test, OpenAI's o3, found about half of the known errors in the papers (when instructed to identify up to five errors), providing further evidence that LLMs can help flag serious issues, though with limitations in scope and sensitivity.

*Advanced systems*

These studies, while encouraging, were conducted using relatively simple implementations of now mostly deprecated models. Recent technical advances suggest substantially greater potential for AI-assistance with different aspects of the peer review process. Modern LLMs incorporate chain-of-thought reasoning capabilities that allow them to break down complex evaluation tasks into sequential analytical steps, leading to improved accuracy and reduced hallucination rates (Wei et al., 2022; Yao et al., 2023). For example, OpenAI's o1 (released in 2024) achieved 74% accuracy on Olympiad-level problems compared to 12% for GPT-4o (OpenAI, 2024). Additionally, tool-use capabilities enable LLMs to access external databases for citation verification, statistical validation, and cross-referencing claims against published literature (Schick et al., 2023). Rather than relying solely on parametric knowledge (that is, the internal representation of knowledge stored in model weights), these systems can access external knowledge, for example via web searches or Application Programming Interface (API) requests. This can greatly reduce hallucinations and increase accuracy (Gao et al., 2023).

The discussion so far has focused on commercially available, off-the-shelf, generic generative AI systems based on LLMs like OpenAI's GPT/o series. However, it is possible to technically adapt general models to a specific function or area. For example, this can be done through further training on a specialized dataset ('fine-tuning') or by giving models access to an external knowledge base ('retrieval-augmented generation' or RAG) (Lewis et al., 2020). And some of the most promising applications of LLMs in peer review might come from such specialized, application-specific language models (cf. Porsdam Mann et al., 2025; Liddicoat et al., 2025). For instance, the OpenReviewer system is an LLM fine-tuned on 79,000 expert reviews from machine learning conferences. The system's outputs matched human review recommendations at more than twice the rate of the best LLM used in the comparison (GPT-4o), while maintaining rating distributions that more closely match human reviewers than general-purpose LLMs (Idahl & Ahmadi, 2024).

Fine-tuning and RAG can also be combined. A journal might fine-tune a model on curated past reviews while giving it RAG access to its style guide, review checklist, scope statement, and recent publications, thus creating a system that evaluates manuscripts against both disciplinary standards (via fine-tuning on past reviews) and journal-specific expectations (via RAG access; on the risk, however, of this introducing a conservative bias against innovative contributions, see below). Lin (2025b) pushes this logic further, proposing that empirical papers themselves be restructured to include machine-readable "structured appendices"—essentially bundled data, code, and metadata that would allow an AI system to re-run (attempt to replicate) the reported analyses rather than simply "reading" the author's description of them. The idea is that for claims amenable to computational verification, the AI reviewer would not need to rely on trust (in the author-given claims) or inference; it could verify directly. This would not, of course, apply to all forms of scholarly argument; but for research that is already quantitative or otherwise computational, it represents a substantial expansion of what AI-assisted review could actually check.

In theory, these approaches enable customization at multiple levels: journals could develop models aligned with their editorial voice and publication standards; fields could create specialized LLM-reviewer tools for methodological approaches unique to their discipline; and individual reviewers could fine-tune (or otherwise train) personal models on their previous reviews regarding similar topics (cf. Porsdam Mann et al., 2023), creating AI assistants that maintain their human-user's analytical style while handling routine evaluation tasks.[1]

More sophisticated architectures promise further enhancements. 'Agentic' systems are LLMs trained to break down overarching goals into multiple component tasks, with the ability to plan and interact with external environments to achieve these tasks autonomously (Park et al., 2023). Multi-agent systems extend this concept by enabling communication between multiple AI agents, with each agent performing one or more specialized roles (Han et al., 2026). The Multi-Agent Review Generator (D'Arcy et al., 2024) demonstrates the potential of this approach in the peer review context. In this setup, different LLM agents are assigned specific functions corresponding to specific sections or features of a paper—i.e., methodology, clarity, or impact—and then programmed to collaborate with each other. D'Arcy et al. (2024) reported that this system produced an average of 3.7 'good' comments (as rated by human researchers in a blinded user study) per paper, compared with 1.7 'good' comments per paper from a single-LLM configuration (they also reported a decrease in generic comments from 60% to 29%).

Similarly, the Generative Agent Reviewers (GAR) framework (Bougie & Watanabe, 2024) combines agents with distinct reviewer personas and a meta-reviewer to synthesize their output. In pairwise comparisons judged by both GPT-4o and human evaluators, GAR's reviews were preferred over those of other LLM review systems and, in both evaluations, also over human reviews, though the margin was narrow in the human evaluation. Beyond performance improvements, these multi-agent systems can be used to experiment with factors influencing peer review. Jin et al. (2024) tested various social science theories (i.e., groupthink, authority effects) in their AgentReview framework, concluding that such social factors accounted for 37.1% of the variation in simulated paper decisions.

[1] A natural concern is that such personalization could calcify a reviewer's existing habits—producing reviews that mechanically recycle the reviewer's familiar criticisms rather than engaging with what is distinctive about the manuscript at hand. This is a real risk, but it is worth noting that it is also a risk of unassisted human review: Reviewers already have characteristic preoccupations and blind spots, and these already show up, paper after paper, without any AI involvement. The question is whether a well-designed system could do better—for instance, by flagging where its draft commentary is generic (i.e., not responsive to specific features of the current manuscript), or by prompting the reviewer to consider dimensions they have historically underweighted. In principle, an AI trained on a reviewer's past outputs is also well-positioned to *model* that reviewer's tendencies, and could be explicitly instructed to compensate for them—something the reviewer alone cannot easily do.

A multi-agent system reviewing an interdisciplinary manuscript could simultaneously evaluate it from multiple perspectives, helping to address the growing challenge of reviewing increasingly interdisciplinary research that often exceeds the expertise of individual human reviewers. Multi-agent systems could also be valuable in disciplines where there might be a lack of consensus on higher-order matters or fundamental theoretical commitments—for example, moral philosophy (e.g., consequentialism vs. deontology), statistics (e.g., Bayesian vs. non-Bayesian), and so on. Such systems might mitigate biases that stem from reviewers' underlying empirical or normative commitments; for example, they could lessen the influence of a libertarian reviewer's predispositions when evaluating a manuscript defending the permissibility of paternalism in public health. The core idea of combining multiple LLMs can be taken even further, through workflow orchestration and ensemble systems (a method predating modern LLMs; Dietterich, 2000) which coordinate specialized models or aggregate the output of multiple models, respectively. While the details are beyond the scope of this discussion, it is important to note that the technical design space for multi-LLM peer review is still largely unexplored.

Generating diverse evaluations does not, of course, ensure they will be weighted fairly, whether by a human or an LLM reviewer. A reviewer who receives both a libertarian, and a utilitarian critique can simply set aside whichever perspective they find uncongenial, just as they might when reading two human reviews that pull in opposite directions. We cannot guarantee that either reviewers or authors will engage seriously with positions they are predisposed to reject. What we can do, however, is to ensure those positions are presented. A practical approach would be a two-step process/framework in which one AI agent reads the manuscript and identifies the relevant evaluative angles, and accordingly, initiates a separate agent for each of those angles. This does not solve the deeper problem of motivated or selective reasoning, but it does mean that alternative viewpoints are at least generated, documented, and made available—something the current human-centred system does not reliably provide.

Taken together, the current body of evidence highlighted above is still small, but the results are striking. At least across the limited domains studied to date and reviewed above, LLMs can approach or exceed human baselines on some proxies of review quality. Given the nascency and heterogeneity of the evidence, blanket claims are unwarranted. A defensible reading is that targeted, supervised application-specific LLM assistance for peer review is plausible and makes it worthwhile to conduct tightly scoped pilots.

## Objections and responses

Even if preliminary findings suggest that LLMs are not inferior in quality to human reviewers, some concerns go beyond technical performance. In what follows, we attempt to set out these objections in their strongest form and consider possible responses. Our aim is not to prescribe specific policies but to assess whether the principal objections to AI-assisted peer review withstand scrutiny when set against the status quo, available evidence, and potential mitigations.

Most of the objections below concern possible *risks* and *benefits* of AI involvement in various aspects of the peer review process. Before turning to those, however, we begin with an objection that there is something inherently problematic about the very notion of an AI "peer" reviewer, because AIs cannot be "peers" to human authors in some relevant sense. In dealing with this objection, we make clear that we are not proposing that LLMs—or any (agentic) AI system—would itself be making final decisions about the publishability of a submitted manuscript; rather, we propose that AI systems could potentially assist with *some* aspects of the peer review process, while final reviewer and editorial judgments would be made by humans, who would also need to take responsibility for these.

## 1. Ontological Confusion

### The Objection

The term "peer review" is not descriptive but normative: it specifies that evaluation should come from one's *peers*—fellow researchers who share relevant expertise, disciplinary commitments, and, crucially, membership in the scholarly community whose norms and standards the review process is meant to uphold. An LLM is not a peer in any of these senses. It holds no appointments, bears no reputational stakes, has no history of contribution to a field, and stands in no relationship of mutual accountability with the author. To call its output "peer review" is therefore a category error, regardless of how closely that output resembles a competent human review in form or content. On this view, the problem is not (only) that AI might perform the task poorly—it is that AI *cannot* perform the task at all, because the task is constitutively defined by the kind of agent who carries it out. A review produced by an entity that cannot be held accountable, that has no stake in the integrity of the field, and that participates in no community of practice is not peer review even if it happens to contain useful observations.

The concern is not only semantic. Peer review functions in part because it embeds evaluation within a web of professional relationships: reviewers are selected because of their standing in a field; their judgments carry weight because they have earned credibility through their own published contributions; and the process works (when it does) because reviewers internalise norms of fairness, constructiveness, and intellectual honesty that are cultivated through participation in a shared enterprise. An LLM cannot internalise norms in this sense. It can be instructed to simulate adherence to them, but simulation and genuine normative commitment are not the same thing. If we allow AI-generated evaluations to shape publication decisions, even indirectly, we risk hollowing out the social and epistemic foundations on which the authority of peer review depends.

### Our Response

This objection raises important conceptual questions, and we largely accept its premise: an LLM is not a peer, and a report generated by an LLM is not, in itself, peer review. But the objection gains its force from a proposal we are not making. At no point in this paper have we argued that AI systems should serve as *the* reviewer of record for a manuscript—that is, as the agent whose judgment determines whether a paper is accepted, revised, or rejected. What we propose is that LLMs can assist human reviewers and editors by producing draft critiques, flagging potential errors, checking references, identifying relevant literature, and organising evaluative observations—all of which must then be accepted, rejected, modified, or supplemented by a human who *does* bear the relational and epistemic standing that the objection rightly emphasizes. The human reviewer or editor remains the peer. The LLM is a tool ~~that~~ the peer uses.

This distinction tracks a difference in how responsibility is assigned and how evaluation is exercised. When a reviewer employs a reference manager to check citations, or a statistical package to verify reported analyses, we do not say that the software has conducted part of the peer review—even though its outputs directly inform the reviewer's judgment. The same logic applies to LLM-generated observations: they become part of the review only insofar as a human expert endorses them, and that expert remains accountable for the final report. What changes is not *who* is reviewing but *what tools the reviewer has access to* while doing so.

That said, the objection does illuminate a genuine risk, which is that the distinction between "tool" and "reviewer" could erode in practice even if it is maintained in principle. If a reviewer submits an LLM-generated report with minimal modification, the human is functioning as the reviewer only in name. We take this concern seriously, and it is addressed in our discussion below. The safeguards proposed there, including training in critical evaluation of AI outputs, and transparency about the extent of AI involvement, are designed to prevent this erosion.

There is, however, a further dimension to this issue that the objection overlooks: the perspective of authors. Current prohibitions on (certain types of) AI use in peer review are typically imposed by journals and directed at reviewers, without consulting the researchers whose work is being evaluated. Yet authors have a legitimate interest in how their manuscripts are assessed. Some authors may reasonably prefer that AI-assisted tools be used if this means faster turnarounds, more comprehensive error-checking, and/or more systematic coverage of relevant literature—particularly if the alternative is a months-long wait for a brief, superficial human review of the kind that the current system, as documented above, regularly produces. Other authors may prefer to receive feedback only from human reviewers, even if this entails longer processing times. (Naddaf, 2025). Both preferences may be reasonable, and both deserve consideration.

A more principled approach would give authors a meaningful choice. At the point of submission, journals could, perhaps, allow authors to indicate whether they consent to AI-assisted evaluation of their manuscript, and if so, under what conditions (e.g., only for specific tasks such as reference verification or statistical checking; only with particular confidentiality safeguards; or for more substantive evaluative assistance). Authors who opt out would be routed to a fully human review process, with the understanding that timelines may differ accordingly. This is analogous to existing options that some journals offer regarding, for example, single-blind versus double-blind review, or the choice to post a preprint—matters where author preferences are recognised as relevant to how the evaluation process should proceed.

Framing AI assistance as a matter of informed consent rather than blanket prohibition respects author autonomy, acknowledges that different researchers may weigh the relevant trade-offs differently, and avoids the paternalism of assuming that all authors share the same views about what constitutes an acceptable review process.

## 2. Hallucinations and Biases

The Objection.

LLMs are known to generate "hallucinations": plausible sounding but false statements, fabricated citations, and spurious critiques disconnected from manuscript content (Donker, 2023; Huang et al., 2025). Such errors are not anomalies but an inherent consequence of how LLMs generate text: models produce outputs by predicting likely next tokens based on statistical patterns in training data, which means they will sometimes generate plausible-sounding content that has no factual basis (Huang et al., 2025). Access to external tools and knowledge sources—such as citation databases, statistical software, or web search—can substantially reduce the frequency of hallucinations by grounding model outputs in verifiable information, but cannot eliminate them entirely. In the context of peer review, where legitimacy depends on accurate representation of empirical findings and logical argument, even a low hallucination rate poses serious risks: an AI-assisted review could introduce specious criticisms, endorse claims the manuscript does not actually support, or cite sources that do not exist—any of which could distort editorial decisions and, if undetected, corrupt the scholarly record.

Also concerning is that LLMs, trained on text corpora, encode and can amplify structural biases from those sources. This includes tendencies to equate quality with institutional prestige, established methodological norms, authors from specific demographic groups or locations, or already well-established authors, potentially disadvantaging novel approaches or individuals from marginalized groups. Such biases are not hypothetical: one large-scale experiment in economics, covering nearly 9,000 submission evaluations, found that LLMs tended to favour prominent institutions, male authors, and renowned economists, despite controlling for manuscript quality (Pataranutaporn et al. 2025)

Our Response

The problems of accuracy, hallucinations, and bias go to the heart of what peer review is supposed to accomplish; that is, safeguarding the integrity of the scholarly record. A quality control system that itself introduces errors and biases is clearly a system that fails at its fundamental purpose. Any use of LLMs in the peer review process must therefore carefully consider both the probability and magnitude of these risks, and the available mitigation strategies.

Yet it is crucial to recognize that many so-called hallucinations occur when LLMs are used "off-the-shelf" and in isolation, without access to the very tools that human reviewers themselves take for granted. Requiring an LLM to review a manuscript without access to: article databases to verify references; a web search to confirm current information; or an "interpreter" to run code and check mathematics, is equivalent to asking a human reviewer to work entirely "from memory"—without a reference library, a browser, or a calculator. At a minimum, LLMs should be given access to the citations and sources they are expected to evaluate. When generic AI is used without access to the necessary tools and knowledge corpus to generate accurate answers, hallucinations are better understood not as intrinsic flaws but as a kind of *user error*: The model generates a plausible example or template when it cannot access the data needed to provide a definitive answer (Earp et al., forthcoming).

Several technical approaches can substantially mitigate hallucination risks, even if they cannot eliminate it entirely. Agentic models equipped with web search, tool integrations, and protocol connections (Model Context Protocol (MCP)/Agent Communication Protocol (ACP)) can concertedly verify citations against scholarly databases, check calculations through computational engines, and access discipline-specific knowledge bases in real time—grounding their outputs in external evidence rather than relying solely on parametric memory. Fine-tuning, RAG, and multi-agent architectures further improve accuracy by anchoring model outputs in curated, domain-specific sources (Idahl & Ahmadi, 2024; D'Arcy et al., 2024). While building such systems requires some degree of AI literacy, no- and low-code solutions (such as n8n or Claude Code) are making them increasingly accessible.

Moreover, we must remember that the current human system already falls far short of its ideals. A fair evaluation of the ethics of LLM use in peer review should compare the current state of the art to the status quo, not yesterday's models to an idealized human reviewer. That status quo, as documented above, is deeply problematic: reviewers detect only a quarter of major errors (Godlee et al., 1998), show inter-reviewer agreement barely above chance (Bornmann et al., 2010), demonstrate systematic biases favoring prestigious institutions and male authors (Murray et al., 2018; Tomkins et al., 2017), and impose delays of months to years (Andersen et al., 2021).

Critically, AI biases may prove more tractable than human ones. Once identified through pre- and post-deployment audits, they can be addressed through retraining, debiasing techniques, or ensemble methods incorporating diverse perspectives. Human biases, rooted in decades of socialization, resist such direct intervention. Moreover, we can audit AI systems for bias in ways impossible with humans, running thousands of controlled experiments varying author characteristics while holding manuscript content constant (cf. the aforementioned AgentReview study, which observed social factors explained roughly a third of variance in peer review; Jin et al. 2024). Lastly, blinding reviewers, whether human or AI, to the identity of the submitting authors can at least partially address this issue for both types of reviewers.

Finally, and perhaps most importantly, we are not proposing that AI systems conduct peer review independently (see Objection 1). Existing standards for LLM use in scholarship clearly mandate that humans using these systems must take full responsibility for the accuracy of any LLM-produced text (cf. Ganjavi et al., 2024; Porsdam Mann et al., 2024; Luo et al., 2025). Similar standards can, and should, be insisted on for LLM use in peer review, a point we return to below (see Objection 4 and associated response).

## 3. Novelty Recognition and Homogenization

The Objection

LLMs predict the next most likely token (a fragment of a word, other symbols, or characters) in a series of tokens, given patterns in their training data (Bender et al., 2021). Thus, even in an idealized case of "perfect" prediction, this mechanism constrains models to reproduce and recombine what is already known. The implication is that systems built to extend past distributions of language may struggle, in principle, to evaluate work that departs radically from established paradigms.

Empirical evidence reinforces this concern. Juzek and Ward (2025) found that 21 specific words, including "delve," "intricate," and "underscore", have increased dramatically in scientific abstracts since the widespread adoption of LLMs. Intriguingly, their investigations point towards reinforcement learning with human feedback (RLHF) as the causal mechanism. RLHF is a technique used to align LLMs with human preferences. When models are fine-tuned to maximize human approval, they apparently learn to favor words and phrases that evaluators find satisfying—typically those that sound authoritative, comprehensive, and familiar. It may be that this preference for the conventional extends beyond vocabulary; if RLHF aligns models to favour linguistic patterns that humans find appealing, it may similarly train them to prefer conventional scientific ideas over radical departures from established thinking. In literature discovery, AI systems can act as powerful gatekeepers that amplify existing biases like the Matthew effect, in which those who already have advantages accumulate further advantages quicker than others (Lin, 2025a); one study found that over 60% of AI-generated paper recommendations were for articles in the top 1% of citation distributions (Algaba et al., 2025). Relatedly, a recent study found that scientists using AI had published more papers and received more citations, but focused on narrower topics and spent less time collaborating with other scientists (Hao et al., 2026).

A further dimension to this issue relates to cultural and linguistic patterns. LLMs perform best with languages and intellectual traditions that are well-represented in their training data (Bender et al., 2021). Research grounded in less-represented philosophical frameworks, regional scholarly traditions, or languages with limited digital corpora may not be adequately evaluated. For instance, an AI system trained primarily on Western academic literature might lack sufficient exposure to evaluate research employing Ubuntu philosophy, Buddhist logic, or other non-Western analytical frameworks. This disparity in representation stems from practical constraints but nonetheless creates systemic disadvantages for scholarship from underrepresented traditions. Such technical limitations risk amplifying existing disparities in whose knowledge gets recognized and validated in the global scientific community.

Our Response

These concerns about innovation and diversity strike at the heart of scientific progress: Science advances in part through paradigm shifts (Kuhn 1962/1970) that violate existing norms, and underrepresented traditions are often a key source of alternative ways of engaging with a problem that might lead to such breakthroughs (Hofstra et al., 2020). Yet, as is the case with hallucination and biases, there are both technical and procedural methods that can at least mitigate these issues, if not solve them entirely.

In practice, contemporary AI systems have demonstrated the capacity to generate novel content, from new mathematical proofs and problem-solving strategies (Romera-Paredes et al., 2024) to original protein structures (Jumper et al., 2021) and chemical compounds (Bran et al., 2024). These outputs suggest that statistical prediction can produce innovation, particularly when models are scaled and equipped with reasoning strategies such as chain-of-thought prompting or tool integration. But here too, it must be borne in mind that human peer review also demonstrates conservative bias against innovative work (Guthrie et al., 2019). The difference lies in how these limitations manifest and might be addressed.

Importantly, LLMs' broad training might sometimes enable them to recognize novelty that specialist reviewers miss. While a reviewer deeply embedded in one paradigm might reject work that violates their field's assumptions, an LLM trained across multiple disciplines (or a multi-agent setup with individual LLMs ~~each~~ singularly trained on a specific discipline) might recognize when concepts from one domain offer revolutionary potential in another. The system might identify that what appears as "error" from one paradigm's perspective represents valid methodology from another. This cross-pollination capacity could help identify certain forms of innovation, even if it turns out that the system cannot generate truly novel paradigms itself.

The homogenization concern, while related to this fundamental limitation, may be simpler to address because it reflects additional layers of design choice. The RLHF processes used in general-purpose LLMs optimize for broad human preferences, such as authority and clarity. While such optimization may serve commercial chat applications, it may ~~also~~ prove problematic for scientific peer review where breakthrough thinking often initially appears incomplete, unclear, or challengingly unconventional. When models are trained to maximize human approval (i.e. their outputs are non-adversarial, and in some sense "sycophantic"), they learn to reproduce what feels satisfying and familiar rather than what might prove revolutionary. This suggests that for peer review applications, technical interventions are perhaps even more crucial than for other concerns.

Application-specific fine-tuning offers a direct counter to these homogenizing pressures. The OpenReviewer system, when trained specifically on peer review data rather than general human preferences, not only achieved superior accuracy but maintained rating distributions matching human reviewers' full range of peer review judgments, avoiding the excessive positivity of general LLMs (Idahl & Ahmadi, 2024).

Procedurally, editorial boards and funding agencies must make deliberate choices about what kind of diversity they wish to preserve. This means recognizing that the apparent trade-off between quality control and innovation may itself be an artifact of how we currently configure these systems. With appropriate design, AI assistance could potentially identify and champion the very innovations that human reviewers, constrained by their training within existing paradigms, might miss.

## 4. Human Oversight, Excessive Deference, and De-skilling

### The Objection

The previous two objections concerned the technical performance of LLM systems. However, there are other fundamental concerns relating not to the models themselves, but rather to their effects on human agency and expertise. It is often said that AI systems should retain a "human-in-the-loop"—that is, that human oversight is crucial to ethical and effective AI use. Yet, if AI systems require human validation, then the validators must possess adequate expertise and spend sufficient time to critically evaluate AI outputs—and if they possess such expertise and time, why do they need AI assistance?

A related concern is the deference problem, otherwise known as automation bias. Multiple studies document that those charged with overseeing AI systems, for example in healthcare applications, gradually surrender their critical judgment, accepting AI recommendations even when their own expertise suggests otherwise (Abdelwanis et al., 2024; Nguyen, 2024). In peer review, such deference could mean that reviewers might accept AI-generated critiques without the careful scrutiny they would apply to their own thoughts, particularly when under time pressure or reviewing outside their core expertise.

Moreover, peer review serves a crucial pedagogical function that excessive AI assistance threatens to erode. Reviewing others' work teaches essential skills: identifying methodological flaws, evaluating evidence quality, offering constructive criticisms, and appreciating diverse research approaches. These skills may then transfer to a scholar's own work, improving study design and presentation. The concern is not restricted to individual skill development but extends to the collective capacity of the scientific community to maintain critical standards. Uncritical delegation can foster AI complacency, a progressive disengagement from the evaluative struggle through which expert judgment is forged (Lin and Sohail, 2026).

### Our Response

Evidence from other fields such as radiology and aviation demonstrates ~~that~~ these concerns about oversight, deference, and de-skilling are indeed serious. In radiology, Dratsch et al. (2023) discovered even highly experienced radiologists exhibited automation bias when using AI assistance, changing correct diagnoses to incorrect ones based on AI suggestions. Research in aviation has demonstrated skilled pilots can lose critical competencies when over-relying on autopilot systems, leading to catastrophic failures when manual intervention becomes necessary (Mosier et al., 1998). Yet this same body of evidence also helps us understand both the extent of these problems and which mitigation strategies prove effective.

What is framed as de-skilling should instead be understood as a necessary recalibration of expertise. Effective AI integration demands the cultivation of essential AI meta-skills: the strategic direction of generative systems toward rigorous analysis, the critical discernment of their outputs against disciplinary standards, and the systematic calibration of human-AI collaboration through iterative refinement (Lin and Sohail, 2026). This follows established patterns of technological augmentation. When calculators became ubiquitous, the frequency of by-hand calculation diminished, but this freed up cognitive capacity for tackling more advanced mathematical and scientific problems (Voinea et al., 2026). Similarly, fifth-generation fighter jets automate many lower-level piloting tasks, enabling pilots to focus on higher-level situational awareness, coordination, and tactical decision-making. The same logic applies to LLMs in peer review: if routine labor such as formatting checks, literature verifications, or drafting generic critiques is delegated to AI, human reviewers can devote more attention to conceptual clarity, methodological innovation, and the broader significance of research findings.

First, however, let us consider the role and nature of oversight. It is true that needing human oversight reduces the efficiency gains one might expect from a fully automated system; a feature of LLM use that Ohde et al. (2025) label a "new and tedious burden". Yet this shift from generation to oversight is neither new nor unique to AI. Academic careers already follow this trajectory: junior researchers write papers, senior researchers review them; PhD students conduct experiments, supervisors evaluate results, and so on (Hurshman et al., 2025). The progression from assistant to full professor involves steadily less content creation and more quality control, a transition we accept as natural professional development rather than problematic burden. Indeed, the peer review system itself rests on the recognition that expert evaluation of others' work provides value even when the evaluator could theoretically have produced similar work themselves. The efficiency gain comes not from eliminating oversight but from the differential effort required: reviewing a manuscript takes hours, writing one takes months; similarly, fact-checking, editing, and generally improving an LLM-generated/-assisted review draft might take a couple of hours, but this is likely less than the handful to dozens of hours needed to write one from scratch.

The deference problem (automation bias) represents perhaps the most serious of these concerns. However, precisely because it is a widespread issue across several important domains, it is also one that has been studied in detail. In general, three types of solutions have been studied, each of which helps to mitigate the problem: (1) training programs that emphasize user accountability for decisions, (2) design features that present confidence levels alongside automated recommendations, and (3) providing information rather than direct recommendations, requiring users to integrate and interpret data themselves (Goddard et al., 2012). Research has also shown that automation bias is strongly correlated with cognitive load (Lyell and Coiera, 2017). These findings could help inform mitigation strategies in the peer review context. Journals, publishers, or universities could provide training courses in the effective use of LLMs in peer review. They could provide prompt libraries or even adapt models (via fine-tuning or custom instructions) to provide relevant information as opposed to direct accept, revise, or reject decisions. They could instruct, advise, or require reviewers to use prompts or models designed to elicit confidence scores rather than direct requests. When an AI system flags that it has low confidence in a particular critique, reviewers know to examine that aspect more carefully.

Importantly, these same mitigation strategies also address the de-skilling concern. Training programs that reduce automation bias by teaching critical evaluation of AI outputs simultaneously preserve and develop the analytical skills that peer review is meant to cultivate. When reviewers must actively integrate information rather than passively accept recommendations, they maintain the cognitive engagement necessary for skill development. Furthermore, the de-skilling concern must be weighed against current realities. A harried postdoc reviewing their tenth manuscript while preparing job applications may gain little pedagogical benefit from rushed, mechanical review. Delegating part of this workload to AI would also help alleviate the significant opportunity costs faced by early career researchers.

Moreover, researchers develop critical evaluation skills through multiple activities beyond peer review: journal clubs, conference presentations, grant panels, and collaborative exchanges. If AI reduces review burden, saved time could be invested in other activities that develop evaluation skills more effectively than the current review system.

We acknowledge that none of this guarantees that reviewers will utilize AI well in practice. Permitting sound use will inevitably produce some unsound use too. But this gap is not unique to AI policies. Many policies whose value are not seriously disputed are observed with varying degrees of strictness. For example, conflict-of-interest disclosures are routinely required yet often perfunctory. In general, imperfect adherence has not been treated as grounds for abolishing the underlying practice. More directly, we do not believe we are in a situation where the downside risk of permitting AI use outweighs the likely benefits. The relevant comparison is not between ideal AI-assisted review and the status quo, but between imperfect AI-assisted review and the imperfect human system documented above. If AI assistance, even imperfectly applied, addresses some of these issues, even partially, then the case for strategic piloting and phased adoption (with ongoing testing) holds.

## 5. Confidentiality

### The Objection

The use of AI systems for peer review raises fundamental concerns about the confidentiality of unpublished research. Many journals and funding agencies currently prohibit AI assistance on the grounds that uploading manuscripts to commercial AI tools/providers could expose novel findings that represent years of intellectual investment (e.g., NIH, 2023). Since priority determines credit in many fields, even minor leaks could prove professionally devastating. A researcher's career advancement, grant funding, and scientific legacy often depend on being first to publish key discoveries.

The concern applies at multiple levels. At the technical level, there are questions about data persistence: once a manuscript enters an AI system, does its content remain accessible in some form (Carlini, 2021)? At the legal level, uncertainty exists about data ownership and usage rights when academic content is processed by commercial entities whose terms of service may claim broad licenses to user inputs (Chesterman, 2025). At the ethical level, submitting authors' unpublished work to AI systems without explicit voluntary consent arguably violates the trust relationship fundamental to peer review (NIH, 2023). At the competitive level, if manuscripts enter training datasets, future users working on similar problems might gain unfair advantages through model behaviors influenced by that training, even without direct data extraction (Earp et al., 2025).

Unlike published papers, manuscripts under review contain ideas at their most vulnerable; polished enough to reveal their value but not yet protected by publication priority. The peer review process requires authors to share these ideas with journals and reviewers based on strict confidentiality assurances. Introducing LLM systems into this process without absolute guarantees of data security could be seen as betraying this foundational trust.

Our Response

While confidentiality concerns deserve serious attention, the current policy landscape is uneven and often overreaches. Publisher policies range from targeted restrictions on uploading manuscripts to commercial AI tools without author consent, to blanket prohibitions on any use of AI in the review process. But the confidentiality rationale applies only to a subset of possible LLM uses. It does not, for instance, apply to locally hosted models, enterprise deployments with contractual data protections, or uses that do not involve sharing manuscript text with external services.

The confidentiality rationale also sits uneasily with existing practice. Manuscripts are routinely stored, transmitted, and processed using web-based submission and review platforms hosted on third-party servers. Editors and reviewers regularly access confidential content via Gmail, Outlook, Microsoft Word Online, Google Docs, and other cloud services. If it is considered acceptable to entrust manuscripts to Google Cloud or Microsoft Azure for email, document preparation, and journal management systems, it is not clear why equivalent assurances of confidentiality from Google Gemini, Microsoft Copilot, or other AI services should be judged categorically differently. Provided that the same enterprise-level contractual protections and compliance standards apply (see e.g. OpenAI, 2025), the risk profile is not fundamentally new.

Major commercial LLM providers such as OpenAI and Anthropic offer explicit options which users can toggle in their settings to opt out of contributing their data to model training. If permitted to use commercial systems, reviewers could and should be made aware of this option, perhaps as a routine aspect of reviewer invitations.

For institutions requiring absolute data control, technical alternatives eliminate external dependencies entirely. Open-source language models can be deployed on local infrastructure, ensuring complete data isolation. While early open-source models lagged commercial offerings significantly, this gap has narrowed dramatically. Models like DeepSeek-R1, Kimi K2, Mistral, and Llama 4 rank highly on various metrics and benchmarks that are used to evaluate model performance (see, e.g. Guo et al., 2025; Moonshot AI, 2025).

One must also keep in mind the comparison to human reviewers, who are known to have delayed manuscripts to publish competing work, appropriated ideas for grant proposals, or shared content without permission with colleagues and students (e.g. Jackson et al., 2026). Human reviewers have career incentives, competitive pressures, and personal biases that can motivate confidentiality breaches. They might unconsciously incorporate ideas into their own thinking (Earp et al., 2025), or deliberately exploit their privileged access. LLM systems, whatever their limitations, have no papers to publish, no grants to win, and no careers to advance. They cannot engage in the deliberate copying or misuse that represents the most serious confidentiality threat in peer review.

This is not to dismiss confidentiality concerns but to evaluate them proportionally. Properly configured AI systems may present no greater confidentiality risk, and potentially lesser, than existing practice. The key is recognizing that confidentiality protection is not binary but involves choosing appropriate tools for specific contexts, just as we currently do elsewhere throughout the research process.

## 6. The Efficiency Objection, or: Opening the Floodgates

### The Objection

Deploying AI-assisted peer review risks exacerbating a longstanding problem in academic publishing: the sheer volume of research getting published. Between 2016 and 2022, the number of indexed articles increased from ~1.92 million to ~2.82 million (Hanson et al., 2024). Many scholars have warned that this growth makes it increasingly difficult to identify and absorb relevant findings (Chu & Evans, 2021; Houssard et al., 2024). The rising number of submissions also places substantial strain on reviewers and editorial systems. While AI could alleviate some of this strain, its efficiency gains might eliminate existing bottlenecks, enabling more articles to pass through peer review more quickly. This trade-off would be especially advantageous for predatory or volume-driven journals, and intensify concerns that quantity is crowding out quality.

### Our Response

Concerns about the rising *volume* of manuscripts are legitimate, but they may be misplaced. In principle, more (good) science is not itself a problem. Indeed, high-quality work should be welcomed regardless of quantity. The real challenge lies in the proliferation of low-quality outputs, especially from predatory or poorly governed journals (Grudniewicz et al., 2019). In such contexts, the worry is not about *too much* science, but about *too much bad science*.

AI-assisted peer review could in fact mitigate this problem by strengthening early filtering: flagging obviously unfit submissions, enforcing basic methodological and reporting standards, and raising the overall baseline quality of manuscripts that reach human reviewers. This could reduce reviewer burden at reputable outlets while making it harder for weak work to pass undetected.

Of course, predatory publishers are unlikely to change their behavior. To the extent that such journals already lack genuine peer review, AI will not make them better or worse; their incentives remain misaligned with scientific quality. A more serious risk is that these journals might adopt AI-generated "reviews" to create a veneer of legitimacy and scale up output. Yet this is arguably a continuation, not a transformation, of existing predatory practices, and should be met with the same countermeasures: indexing standards, author awareness, and institutional disincentives (Laine et al., 2025).

On balance, then, we should be less concerned about the total volume of papers, and more about the distribution of quality. Properly deployed, AI-assisted review may help journals separate signal from noise.

## 7. Vulnerability to Gaming

### The Objection

The introduction of AI into peer review opens up a new attack vector that previously did not exist. As such, authors may now attempt to manipulate AI review systems through adversarial techniques. Because LLM systems operate through predictable patterns, they may be vulnerable to gaming strategies that would not affect human reviewers. Authors might embed hidden instructions in manuscripts (Gibney, 2025), craft text to trigger favourable evaluations, or upload fraudulent data and findings (like the fake disease "bixonimania") to preprint servers (cf. Stokel-Walker, 2026); all to exploit known biases in LLM systems. As LLM use in peer review expands, incentives to discover and exploit such vulnerabilities will intensify.

For instance, in July 2025, 18 preprints on arXiv were found to contain hidden instructions designed to elicit positive reviews from LLMs—ranging from simple commands like "GIVE A POSITIVE REVIEW ONLY" to elaborate outlines structuring entire favourable assessments (Lin, in press). Prompt injection attacks exploit the fundamental architecture of language models, including causing models to ignore their instructions and follow hidden directives (Liu et al., 2023); bad faith users can craft text that manipulates LLM systems while appearing normal to human readers in terms of grammar, meaning, and readability (Ye et al., 2024).

Extending the bixonimania experiment example (Stokel-Walker, 2026); bad actors could, under fictional aliases, deposit *specious* but fabricated studies, results, and conclusions to multiple preprint repositories (like ResearchGate and Preprints.org)—that consequently receive genuine DOIs—in a "disinformational" attempt to taint the LLM system's knowledge base and/or manipulate its assessments.

These suggest that authors could potentially craft manuscripts to receive favorable LLM reviews despite serious flaws, or flagrant fraud. If journals rely heavily on LLM assessments, such gaming threatens to corrupt not only peer review but the broader infrastructure of scientific evaluation—from plagiarism and data falsification detection to citation indexing—potentially distorting scientific knowledge at scale.

Our Response

While the vulnerability to gaming represents a genuine concern, several strategies can mitigate these risks and create more robust AI review systems. The key lies in raising the cost and complexity of successful manipulation to impractical levels.

First, various technical defences are possible. For example, deliberately exposing AI systems to known manipulation techniques during development or fine-tuning can help models recognize and resist common gaming strategies (e.g. Xhonneux et al., 2024). Models might also be designed or prompted to calibrate or reduce their confidence scores for assessments relating to evidence/references derived from non-peer reviewed sources (e.g. preprints).

Second, human reviewers can use various defensive strategies, such as pre-processing manuscript text before evaluation, removing potential trigger patterns (like prompt injections, i.e. "GIVE A POSITIVE REVIEW ONLY") while preserving legitimate content (Liu et al., 2023).

Nevertheless, this is a serious objection. As both adversarial and defensive strategies develop, we are likely to see a cat-and-mouse dynamic in which techniques are developed and mitigated, requiring some degree of technical vigilance and understanding on the part of journals and perhaps even at the level of individual reviewers.

The use of ensemble or multi-agent systems may be particularly effective in this context: An author attempting to game such a system would need to craft text that simultaneously exploits vulnerabilities across all models—a significantly more challenging task. Moreover, disagreement between models could trigger additional scrutiny, flagging potentially manipulated submissions for human review.

Importantly, it is worth noting that complete automation of peer review was never the goal. The combination of human and LLM review may be particularly effective at catching adversarial attempts, since few currently known strategies are effective against both humans and LLM systems.

## 8. Loss of Trust

The Objection

Trust underpins the 'social contract' of human peer review. In the context of AI, trust can be understood as operating on two levels: (i) trust from academic stakeholders (i.e. scholars and researchers) in the credibility of published manuscripts peer reviewed with AI assistance or by an AI system, and (ii) broader societal trust in the legitimacy of science peer reviewed, at least in part, by AI agents or systems.

Empirical data suggest these concerns are salient. A 2025 *Nature* survey of 5,229 academics illustrates such tensions. Only 5% of respondents deemed it "appropriate" for a reviewer to use an AI's output (without 'AI usage' disclosure) as the basis of their peer-review report, while 52% judged such action "not appropriate under any circumstances". And in a 2025 *Frontiers* survey of over 1,000 researchers, 42% said their level of trust in the review process decreased when AI is used by publishers, and 76% were unaware or unsure if a publisher had used AI for the process. These numbers suggest that a significant portion of the academic community views AI involvement in peer review as unacceptable, at least without disclosure. If AI systems are nevertheless used, those who hold such views might lose some degree of trust in the integrity of the scientific record.

Although no studies (to our knowledge) have yet examined societal attitudes towards AI in peer review, it is plausible that public perspectives could be as, if not more, critical as those of researchers.

In sum, AI in peer review risks undermining or eroding the trust that is foundational to the scientific enterprise and which peer review is meant to establish and maintain.

Our Response

Concerns about AI's potential impacts on trust are significant. Yet there are ways to mitigate these concerns substantially. Transparency is perhaps the most fundamental requirement for maintaining trust in AI-augmented peer review processes. This necessitates the adoption of clear, publicly available policies and procedures that govern the use of AI in peer review, whether in initial desk screenings or as support for reviewers. Journals (with their AI developers or providers) should provide technical details of any AI systems they propose to, or actually do, employ. They should also provide information about how these tools would be audited, as well as the results of any audits carried out. Furthermore, journals must provide explicit disclosure regarding when and how AI was employed in the review process. This approach mirrors existing guidelines that require authors to disclose and acknowledge their use of AI in research and manuscript preparation, as similar principles of transparency apply across all stages of the scientific publication process (cf. Porsdam Mann et al., 2024).

Accountability mechanisms must also be clearly established within journal policies to specify responsibility for AI use and potential misuse. When peer reviewers are granted autonomy to use journal-designated AI systems, they must assume full responsibility for the accuracy, integrity, and fairness of their reports. This means that regardless of AI assistance or contribution, the human reviewer remains ultimately accountable for the content and quality of their peer-review report. Such accountability structures ensure that AI remains a tool rather than a replacement for human judgment and expertise. Similarly, journals and publishers should explicitly assume responsibility for the accuracy and reliability of any in-house or licensed systems used for this purpose.

Publishers and journals, including institutions, should also promote literacy about AI capabilities and limitations among all stakeholders in the review process. Authors and reviewers need to be informed through submission guidelines and peer review policies about the journal's designated scope for AI use, including its specific roles and any limitations or restrictions on its application in the peer review process, as well as the rationale for the introduction of AI systems. This educational approach helps ensure that all participants understand both the potential benefits and inherent limitations of AI assistance.

Taken together, these measures can help mitigate concerns about trust and credibility, thereby enabling journals to adopt AI for peer review responsibly while maintaining confidence among both scholars and the broader public.

## Conclusion

The current evidence is preliminary and heterogeneous; yet in several small-scale studies, even retired models like GPT-4 have approached human baselines on specific review tasks. More recent technical advances, including chain-of-thought reasoning, tool use, fine-tuning, RAG systems, and multi-agent architectures, indicate that performance will likely improve dramatically as these methods diffuse into general use. While reasonable people may disagree, we find none of the stated objections fatal to the case for LLM-assisted peer review. The concerns reviewed above are genuine but admit of technical and procedural mitigations that can reduce risks to acceptable levels. Indeed, peer review may represent a particularly promising domain for carefully scoped LLM assistance: some constituent sub-tasks are structured and measurable, the potential benefits are substantial, and AI tools might finally address documented failures that have resisted decades of reform.

The scale and pace of adoption should be proportionate to the evidence. Where carefully scoped pilots show clear gains on pre-defined success metrics, such as shorter turnaround, better error detection, or reduced reviewer burden, the appropriate response is to build on those results by implementing the methods trialed out in pilot form more broadly. Conversely, modest or ambiguous findings warrant further studies rather than immediate implementation. Between pilot and broad adoption lies a range of intermediate steps: phased rollout beginning with lower-stakes tasks, knowledge-sharing between publishers/journals so that lessons are not siloed, and mechanisms for gathering feedback from authors, reviewers, and editors so that problems surface early. Because the demands of reviewing a clinical trial differ sharply from those of reviewing a work in political theory, much of this experimentation will necessarily be field-specific and decentralised; no single policy will fit all disciplines. What matters most is that the process does not stall indefinitely at the pilot stage. The opportunity costs of inaction (the delays, reviewer burnout, preventable errors, and biases documented above) are themselves an ethical problem, and one the current debate too often treats as though it were costless.

The path forward requires neither uncritical enthusiasm nor reflexive rejection, but careful empirical investigation of what works, for whom, and under what conditions. As LLM and general AI capabilities continue to evolve rapidly, so too must our frameworks for evaluating their appropriate use in scientific knowledge production. We invite the research community to engage with these questions through both theoretical analysis and practical experimentation. The future of peer review, and by extension, the validation of scientific knowledge, may depend on getting this balance right.

**Acknowledgements**

In the preparation of this manuscript, GPT-5 and Claude 4.1 Opus were used to edit and rephrase text supplied by the authors. Use of generative AI in this manuscript adheres to ethical guidelines for use and acknowledgment of generative AI in academic research (Hosseini et al., 2025; Porsdam Mann et al., 2024). Each author has made a substantial contribution to the work, has thoroughly vetted it for accuracy, and assumes responsibility for the integrity of their contributions.

**Funding**

This research project is supported by National University of Singapore under the NUS Start-Up grant (NUHSRO/2022/078/Startup/13). This research is supported by NUSMed and ODPRT (NUHSRO/2024/035/Startup/04) for the project "Experimental Philosophical Bioethics and Relational Moral Psychology" with BDE as PI. This research is supported by the National Research Foundation, Singapore under its AI Singapore Programme (AISG Award No: AISG3-GV-2023-012). Zhicheng Lin is supported by Yonsei University Research Fund (5355661) and the Yonsei Fellowship (funded by Lee Youn Jae). TM and SPM were supported in part by a Novo Nordisk Foundation grant for a scientifically independent International Collaborative Bioscience Innovation & Law Programme (Inter-CeBIL Programme, grant NNF23SA0087056).

**Declaration of Interests**

JS is a Bioethics Committee consultant for Bayer.